\documentclass[twocolumn,pra,showpacs,amsmath,amssymb]{revtex4}
\usepackage{amsfonts,epsfig}
\usepackage{amsmath}
\usepackage{amssymb}
\usepackage{amsfonts}
\usepackage{amsthm}
\usepackage{mathrsfs}
\usepackage{color,verbatim,graphics}
\usepackage{url}

\usepackage{graphicx}
\def\be{\begin{equation}}
\def\ee{\end{equation}}
\def\bea{\begin{eqnarray}}
\def\eea{\end{eqnarray}}
\def\bma{\begin{mathletters}}
\def\ema{\end{mathletters}}

\def\0{\overline{0}}
\def\tr{\mbox{tr}}

\def\q0{\underline{0}}

\def\H{{\cal H}}

\def\C{{\mathbb C}}
\def\id{{\mathbb I}}

\def\H{{\cal H}}

\def\N{\mathbb{N}}

\def\tr{\mbox{tr}}
\def\one{\leavevmode\hbox{\small1\normalsize\kern-.33em1}}

\def\bra#1{\langle#1|} \def\ket#1{|#1\rangle}
\def\braket#1#2{\langle#1|#2\rangle}

\def\proj#1{\ket{#1}\!\bra{#1}}

\newtheorem{theo}{Theorem}

\newtheorem{remark}{Remark}

\def\id{{\mathbb I}}
\def\tr{\mbox{tr}}

\begin{document}
%\title{Characterization of quantum correlations with local dimension constraints and its applications}

\title{Characterization of quantum correlations with local dimension constraints and its device-independent applications}
%\title{Characterizing quantum correlations under local dimension constraints and related applications}

\author{Miguel Navascu\'es$^1$, Gonzalo de la Torre$^2$ and Tam\'as V\'ertesi$^3$}
\address{$^1$H.H. Wills Physics Laboratory, University of Bristol, Tyndall Avenue, Bristol, BS8 1TL, United Kingdom\\$^2$ICFO-Institut de Ci\`encies Fot\`oniques, Av. Carl Friedrich Gauss 3, E-08860 Castelldefels, Barcelona, Spain\\$^3$Institute for Nuclear Research, Hungarian Academy of Sciences, H-4001 Debrecen,
P.O. Box 51, Hungary}

\begin{abstract}
The future progress of semi-device independent quantum information science depends crucially on our ability to bound the strength of the nonlocal correlations achievable with finite dimensional quantum resources. In this work, we characterize quantum nonlocality under local dimension constraints via a complete hierarchy of semidefinite programming relaxations. In the bipartite case, we find that the first level of the hierarchy returns non-trivial bounds in all cases considered, allowing to study nonlocality scenarios with four measurement settings on one side and twelve (12) on the other in a normal desktop. In the tripartite case, we apply the hierarchy to derive a Bell-type inequality that can only be violated when each of the three parties has local dimension greater than two, hence certifying three-dimensional tripartite entanglement in a device independent way. Finally, we show how the new method can be trivially modified to detect non-separable measurements in two-qubit scenarios.
\end{abstract}

\maketitle

\section{Introduction}
The realization by John Bell in his 1964 seminal paper \cite{bell} that the correlations arising from measuring space-like separated quantum systems (quantum correlations) can be nonlocal represents one of the most outstanding discoveries of modern physics. The signature of nonlocality, the violation of a Bell inequality, has been extensively verified experimentally and stands as a well established experimental fact \cite{review_exp, review_exp2}.

Besides its foundational interest, quantum nonlocality is instrumental in the emergent field of device independent quantum information processing, whose objective is to infer properties of the underlying state and measurements without assuming any a priori knowledge of the inner working of the devices used. Quantum key distribution \cite{qkd1,qkd2,qkd3,qkd4,qkd5}, randomness generation \cite{random1, random2, random3} and genuine multipartite entanglement certification \cite{genuine, genuine_exp} are celebrated instances of information-theoretic tasks which can be implemented in a black box scenario. The characterization of quantum nonlocality provided by the Navascu\'es-Pironio-Ac\'in (NPA) hierarchy \cite{NPA} played a pivotal role in assessing the security of many of such protocols.

In the last years, it has been pointed out that in many physical situations, e.g., in ion-trap experiments, there is a bound or a promise on the dimensionality of the system under study. Exploiting this promise has led to \emph{semi-device independent} bounds on entanglement \cite{LVB} and novel quantum key distribution \cite{semi_qkd} and randomness generation \cite{semi_rand} protocols more robust and efficient than their fully device-independent counterparts. This approach to quantum information science stems from prior research on dimension witnesses \cite{d_wit1,d_wit2,d_wit3,d_wit4,d_wit5}, which are device independent lower bounds on the Schmidt rank of the bipartite state giving rise to the observed correlations. Clearly, in order to certify the security of semi-device independent communication protocols, or the existence of high-dimensional entanglement in a device independent way, a characterization of quantum correlations under local dimension constraints is needed. In this respect, see-saw variational techniques have proven very useful to characterize such a set of correlations from the inside \cite{seesaw1, seesaw}.

Characterizations from the outside -i.e., the characterization of \emph{limits}- are, on the contrary, problematic. A brute-force approach, advocated in \cite{eisert}, is to reduce the computation of Tsirelson bounds to the minimization of a multivariate polynomial over a region defined by polynomial constraints and run the Lasserre-Parrilo hierarchy of semidefinite programming relaxations \cite{lasserre,parrilo}. Unfortunately, the vast amount of free variables needed to model the simplest nonlocality scenarios makes this scheme intractable in normal computers. Another possibility is to make use of the interesting algorithm proposed by Moroder \emph{et al.} \cite{Moroder}. This method works by implementing a modified version of the NPA hierarchy with extra positivity constraints which effectively bound the negativity \cite{negat} of the underlying quantum state $\ket{\psi}$. By restricting the negativity to be below the value $1/2$, this tool was successfully employed in \cite{Moroder} to derive the maximum violations of the I3322 and I2233 inequalities \cite{I3322, I2233} attainable with qubit systems. In principle, this method can be improved by imposing that not only the state $\psi$ has negativity smaller than $1/2$, but also suitable local postselections of the form $P_AP_B\ket{\psi}$, where $P_A$ ($P_B$) denotes a polynomial of Alice's (Bob's) measurement operators. It is not clear, though, that even this modified scheme converges to the desired set of correlations: indeed, if Peres' conjecture turns out to be false \cite{peres}, there could exist high dimensional states with positive partial transpose which nevertheless produce correlations impossible to reproduce with, say, two qubits.

In the present work, we introduce practical numerical techniques for the full characterization of quantum correlations in scenarios where the local dimension of some parts of a multipartite quantum system are trusted to be bounded, while the local dimensions of the rest of the parties stay fully unconstrained. By exploiting a previously unnoticed connection with the separability problem, we show how to use tools from entanglement detection to characterize the strength of bipartite quantum correlations under local dimension constraints via hierarchies of semidefinite programming relaxations. Combined with the formalism of moment matrices from the NPA hierarchy, the resulting method becomes capable to deal with multipartite scenarios where a subset of the $N$ parties has access to infinite dimensional degrees of freedom. In both cases, the convergence of our sequence of relaxations to the appropriate set of correlations is rigorously proven.

The application of these techniques to several device independent problems is also studied. We use our method to bound the maximal violation attainable via measurements on two-qubit states of a number of bipartite Bell inequalities. This question arises naturally in quantum information science \cite{d_wit1,d_wit2} and convex optimization theory \cite{jop}, where high performance algorithms to solve the problem  are still missing. In addition, we use our tools to derive a tripartite Bell inequality that allows to certify three-dimensional tripartite entanglement in a device independent way, thereby extending Huber \& de Vicente's recent work on multidimensional entanglement \cite{marcus} to the black box realm. We conclude with a semi-device independent application: in \cite{VN} it was proposed a scheme to certify entangling dichotomic measurements under the assumption that the probed states are pairs of independent qubits. We make use our new numerical tools to prove that such a scheme works, i.e., that the linear witness presented in \cite{VN} does actually discriminate separable from entangled measurement operators.

\section{Bipartite non-locality in finite dimensions}
\label{bip}

Consider two separate parties, Alice and Bob, interacting with a two-qubit system with measurement devices which allow them to implement $m$ different dichotomic measurements. Denoting Alice's and Bob's measurement settings by $x$ and $y$ and their measurement outputs by $a$ and $b$, we hence have that $x,y$ range from $1$ to $m$, while $a,b$ can take values in $\{0,1\}$.

Call $Q(\C^2)$ the set of nonlocal probability distributions $P(a,b|x,y)$ which Alice and Bob can generate with this setting. The physical significance of $Q(\C^2)$ is quite clear: if we have the promise that the form of Alice and Bob's state and measurement operators does not vary during the course of the experiment, we should expect to observe distributions in $Q(\C^2)$. A more realistic model, though, would contemplate the possibility that each physical realization of the experiment is different from the previous one, perhaps even depending on Alice and Bob's past measurement history. From the work of \cite{zhang}, we know that in such scenarios any linear Bell-type inequality can be translated into a fully device independent claim: in our case, a claim on the dimensionality of the degrees of freedom Alice and Bob have access to. We will thus be concerned with the problem of conducting linear optimizations over $Q(\C^2)$, or, equivalently, characterizing the convex hull of this set.

Since we are speaking about dichotomic measurements, the extreme points of $Q(\C^2)$ are generated by conducting local projective measurements over a two-qubit state. Let us for the moment restrict to extreme points where all such measurements are rank-one projectors (degenerate cases can be treated in a similar manner), i.e.,

\be
P(a,b|x,y)=\tr\{\rho_{AB}(\Pi^x_a\otimes\bar{\Pi}^y_b)\},
\ee

\noindent where $\Pi^x_a=a\id_2+(-1)^a\proj{u^x}$, $\bar{\Pi}^y_b=b\id_2+(-1)^b\proj{v^y}$, and $\ket{u^x},\ket{v^y}\in \C^2$ are normalized vectors.

We will now show that there is an equivalent way of writing $P(a,b|x,y)$ which will turn out to be very useful. For that, we will map Alice and Bob's state and measurement operators to a $(2m+2)$-qubit state $W$ living in the Hilbert space $ABA_1\cdots A_m B_1\cdots B_m$. The two-qubit space $AB$ will store the shared quantum state $\rho_{AB}$, while the Hilbert spaces $A_1\cdots A_m$ ($B_1\cdots B_m$) will hold Alice's (Bob's) measurement projectors $\{\proj{u^x}\}_{x=1}^m$ ($\{\proj{v^y}\}_{y=1}^m$). The state $W$ is thus given by

\be
W=\rho_{AB}\otimes\bigotimes_{x=1}^m\proj{u^x}\otimes\bigotimes_{y=1}^m\proj{v^x}.
\ee

We can now write

\be
P(a,b|x,y)=\tr(W M^x_a\otimes N^y_b),
\ee

\noindent where

\begin{eqnarray}
&M^x_a=(a\id_{AA_x}+(-1)^a V(A,A_x) )\otimes \id_{A_1\cdots A_{x-1}A_{x+1}\cdots A_m} \nonumber\\
&N^y_b=(a\id_{BB_y}+(-1)^b V(B,B_y) )\otimes \id_{B_1\cdots B_{y-1}B_{y+1}\cdots B_m} \nonumber \\
&
\label{pseudo_meas}
\end{eqnarray}

\noindent Here $V(C,D)$ denotes the SWAP operator between the Hilbert spaces $C,D$. Note that $W$ is a normalized quantum state, fully separable with respect to the partition $AB|A_1|\cdots|A_m|B_1|\cdots|B_m|$.

Conversely, it is easy to see that the convex hull of the set of all distributions $P(a,b|x,y)$ achievable by conducting rank-one measurements over a two-qubit state is given by all $P(a,b|x,y)=\tr(W M^x_a\otimes N^y_b)$, with $W$ fully separable.
%,  $\{B^{xy}_{ab}\}$

Consequently, finding the maximal violation of any Bell inequality $I=\sum_{a,b,x,y} B^{xy}_{ab} P(a,b|x,y)$  in the above systems is equivalent to solve the problem

\begin{eqnarray}
&&\max \tr(W\cdot\sum_{a,b,x,y}B^{xy}_{ab}M^x_a\otimes N^y_b),\nonumber\\
\mbox{s.t. }&&\tr(W)=1, W \geq 0 \nonumber\\
&&W, \mbox{ separable}.
\label{separable}
\end{eqnarray}

Unfortunately, optimizing linearly over the set of separable states is an NP-hard problem \cite{gurvits, gharibian}. Consider then the corresponding Positive Partial Transpose (PPT) \cite{P96} relaxation

\begin{eqnarray}
&&\max \tr(W\cdot\sum_{a,b,x,y}B^{xy}_{ab}M^x_a\otimes N^y_b),\nonumber\\
\mbox{s.t. } &&\tr(W)=1, W\geq 0, \nonumber\\
&&W^{T_P}\geq 0, \mbox{ for all bipartitions }P,
\label{PPT}
\end{eqnarray}

\noindent where $W^{T_P}$ denotes the partial transpose of matrix $W$ with respect to the systems $P$. Note that this condition is a relaxation of the tensor product form of the separable state constraint of the previous problem.

The above problem can be cast as a semidefinite program, and its solution will provide an upper bound on the violation of the said inequality. Note also that we can fix $\ket{u^m}=\ket{v^m}=\ket{0}$, and so, by modifying appropriately the definition of the operators $M^x_a,N^y_b$, we `only' need to optimize over a $2+2(m-1)=2m$-qubit state.

Let us see how this works in practice: take the I3322 inequality \cite{I3322}, a Bell inequality for bipartite nonlocality scenarios with 3 measurement settings and dichotomic outcomes. It follows that we need to optimize over six-qubit PPT states. By imposing the PPT condition over sufficiently many bipartitions, we found the value 0.25, which is known to be achievable via equatorial measurements of the two-qubit maximally entangled state \cite{I3322}, hence obtaining the exact maximal violation.

In general, though, we should expect this method not to return the exact solution. This leads us to consider tighter relaxations of the separability condition. We chose to use the Doherty-Parrilo-Spedalieri (DPS) hierarchy of semidefinite programs to characterize the quantum correlations \cite{trisep}. The intuition behind the DPS method is the observation that any fully separable state

\be
\rho_{1,2,...}=\sum_kp_k\proj{u^1_k}\otimes\proj{u^2_k}\otimes...\otimes\proj{u^n_k}
\ee

\noindent admits an $N$ extension per site of the form

\be
\sigma\equiv \sum_kp_k\proj{u^1_k}^{\otimes N_k}\otimes...\otimes\proj{u^{n-1}_k}^{\otimes N}\otimes\proj{u^n_k}.
\ee

\noindent for any $N\geq 2$. Note that we are not extending the last subsystem. The new state $\sigma$ has the following properties:

\begin{enumerate}
\item
It lives in the Hilbert space $\bigotimes_{k=1}^{n-1}\H_{d_k}^{N_k}\otimes \H_n$, where $\H_d^N$ denotes the $N$-symmetric space of $\C^d$.

\item
It satisfies $\tr_{1^{N_1-1}2^{N_2-1}...}(\sigma)=\rho$.

\item
It is PPT with respect to all bipartitions.
\end{enumerate}

A necessary condition for $\rho$ to be fully separable is thus that a state with the above properties exists. It is easy to see that checking for the existence of such a state can be cast as a semidefinite program. Furthermore, in \cite{trisep} it is proven that the resulting entanglement criteria is complete, even when the last condition is omitted.

\begin{figure}
  \centering
  \includegraphics[width=8.5 cm]{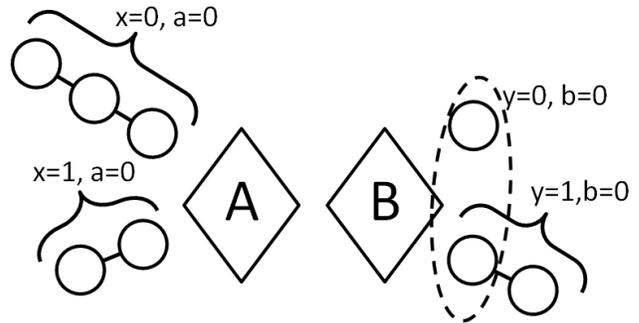}
  \caption{Pictorial representation of the state constraints: the diamonds (heads) denote Alice and Bob state spaces; the circles represent rank-one projectors. $N$ circles joined by a line (legs) must be understood as living in the symmetric space of $N$ particles. In the figure, Alice's projectors $\proj{u^1}$, $\proj{u^2}$ are represented by legs of length 3, 2 respectively. Also, it has been imposed positivity under the partial transpose of one of the circles of Bob's first leg and Bob's second leg.}
  \label{cabezas_patas}
\end{figure}

Coming back to the problem of characterizing non-local correlations in multi-qubit systems, a tighter relaxation for problem (\ref{separable}) is to demand the existence of a state of the form denoted in Figure \ref{cabezas_patas}.

There Alice and Bob's quantum systems are represented by diamonds; we will call such systems \emph{heads}. Circles joined by a line will be called \emph{legs}; they represent Alice and Bob's rank-one projectors. The number of circles in a leg will be the \emph{length} of the leg. Mathematically, a diagram like the above represents a particular relaxation of the separability condition. Namely, drawing a leg of length $N$ for a particular measurement will indicate that we will be approximating the associated rank-one projector by a subsystem of an $N$-symmetric ensemble. In principle, one can demand the positivity of the whole state under the partial transposition of any number of circles (not necessarily belonging to the same leg). Such a condition will be denoted by encompassing with a dashed line the relevant circles.

This time, the action of the operators $M^x_a$ ($N^y_b$) can only be non-trivial in one of the circles of leg $x$ ($y$) and Alice's (Bob's) head. By \cite{trisep}, as we increase the length of the legs on the diagram, we converge to the solution of problem (\ref{separable}).

Informally speaking, the idea behind the method so far is to force the tensor product structure between state and measurements through the $N$ extension of the measurements (legs) and the positive partial transpose conditions. The longer the length, the closer to a tensor product.

\subsection{Higher dimensions and higher number of outcomes}
\label{hig}
In order to optimize dichotomic Bell inequalities over higher dimensional Hilbert spaces, again we can assume that measurements are projective. This time, though, there may exist non-trivial projectors with rank greater than 1. To model a rank-2 projector, we must then introduce two legs, one for each rank-1 projector, and then enforce orthogonality relations between them. Denoting by $C$, $D$ the circles of two different legs, the orthogonality condition is translated as

\be
\tr_{CD}\{W\cdot V(C,D)\}=0.
\ee

\noindent Similar considerations apply to optimizations involving $d$-valued projective measurements.

As for the simulation of generalized measurements with more than two outcomes, note that any POVM can be viewed as a projective measurement in a larger Hilbert space. Namely, for any set of POVM elements $\{M_a\}_{a=0}^{A-1}\subset B(\C^d)$, there exists a complete set of projectors $\{\Pi_a\}_{a=0}^{A-1}\subset B(\C^{d'}\oplus\C^{d})$, such that $M_a=(0_{d'}\oplus\id_d)\Pi_a(0_{d'}\oplus\id_d)$ for $a=0,...,d-1$. Hence, in order to play with $d$-outcome generalized measurements it suffices to consider projective measurements in a larger Hilbert space, project them into the original space and collapse them with Alice's or Bob's head, depending on the case. The amount of resources needed increases very quickly with the number of measurement outcomes, though.

\section{Towards more efficient algorithms and hybrid infinite-finite dimensional optimization}
\label{tow}

As we saw in the last section, with the previous approach, when we enforce the PPT condition (and thus are bound to use SDP), even Bell optimizations in simple scenarios like the $4422$ are intractable with a normal desktop. In this section we will improve the previous algorithm to deal with scenarios where just one of the parties has many measurement outcomes. As we will see, the new algorithm can be extended straightforwardly to deal with multipartite situations where the local dimensions of a subset of the parties are constrained, while the rest have access to infinite dimensional degrees of freedom.

The key to this improvement is a process that we will denominate `body expansion'.

\subsection{Body expansion}
\label{bod}

For simplicity, picture a tripartite scenario where one of the parties, Alice, has total control over her $d$-dimensional quantum system living in the Hilbert space $A$, but we completely ignore the operations being carried out by the other two observers, call them Bob and Charlie, in their Hilbert spaces $B$ and $C$. That is, we are contemplating a nonlocality scenario where the measured correlations are of the form

\be
P(a,b,c|x,y,z)=\tr(\rho_{ABC}\Pi^x_a\otimes E^y_b\otimes F^z_c),
\label{original}
\ee

\noindent where $\{\Pi^x_a\}\subset B(\C^d)$, acting in $A$, are known measurement operators, and $\{E^y_b,F^z_c\}$, acting in $B$ and $C$ respectively, represent unknown projector operators acting over arbitrary Hilbert spaces.

Based on the local mapping approach introduced by Moroder \emph{et al.} \cite{Moroder}, Pusey \cite{pusey} recently proposed to characterize this class of systems by expanding the unknown degrees of freedom in a moment matrix \emph{\`a la NPA} \cite{NPA} while keeping the trusted system the same. This notion can also be found in prior work by Helton \& McCullough \cite{helton}, but, for didactical purposes, we will follow Moroder \emph{et al.}/Pusey's presentation.

Given the multipartite state $\rho$, the idea is to implement the map

\be
\rho\to \tr_B(\id_A\otimes \Lambda_{BC})\rho(\id_A\otimes \Lambda_{BC})^\dagger,
\ee

\noindent with

\be
\Lambda_{BC}=\sum_{|s|\leq n}  s\otimes\ket{s}.
\ee

\noindent Here the sum is over all sequences $s$ of unknown projectors $\{E^y_b,F^z_c\}$ of length $|s|$ smaller than or equal to $n$ (including the identity), and $\{\ket{s}\}$ is an orthonormal basis where each vector is labeled by a sequence of $\{E^y_b,F^z_c\}$. Defining $c^{k,j}_s\equiv\tr\{(\ket{j}\bra{k}\otimes s)\rho\}$, it can be seen that the result of such a map is a positive semidefinite operator of the form

\be
\Gamma^{(n)}\equiv\sum_{k,j}\ket{k}\bra{j}_A\otimes \sum_{|s|,|t|\leq n} c^{k,j}_{t^\dagger s}\ket{s}\bra{t},
\label{expansion}
\ee

\noindent with

\be
\sum_k c^{k,k}_{\id}=1.
\label{normal}
\ee

\noindent From now on, the matrix $\Gamma^{(n)}$ will be called a \emph{generalized moment matrix}. It is worth noting that here we are identifying sequences of operators modulo commutation relations, i.e., $E^y_bF^z_c$ and $F^z_c E^y_b$ are regarded as the same sequence. Also, `null sequences' like $E^y_bE^y_{b'}$, with $b\not=b'$, are not considered, or, equivalently, their corresponding coefficients $c^{k,j}_s$ are set to zero.

If we use (\ref{expansion}) rather than (\ref{original}) to represent the quantum systems involved in the experiment, we will say that the body of parties $B,C$ has been \emph{expanded}. The original probability distribution can be retrieved by

\be
P(a,b,c|x,y,z)=\tr\{\Gamma^{(n)}(\Pi^x_a\otimes \ket{t}\bra{s})\},
\label{proba_ex}
\ee

\noindent where $t,s$ are any two sequences such that $|s|,|t|\leq n$ and $t^\dagger s=E^y_bF^z_c$. It is straightforward to extend the notion of body expansion to more than two parties.

Actually, due to the linear dependence $E^y_{\tilde{b}}=\id-\sum_{b\not=\tilde{b}}E^y_b$, it is enough to consider sequences of projector operators corresponding to the first $A-1$ outcomes in eq. (\ref{expansion}). This allows saving computer memory and leads to the same numerical results, so from now on we will be assuming that generalized moment matrices are only defined on such sequences.

In general, demanding the existence of a positive semidefinite operator $\Gamma^{(n)}$ of the form (\ref{expansion}) constitutes a relaxation of the original problem of characterizing the convex hull of all distributions of the form (\ref{original}). Hence, in order to achieve convergence, we must consider a hierarchy of semidefinite programs $\Gamma^{(1)}\geq 0,\Gamma^{(2)}\geq 0,...$, see \cite{pusey,helton}. However, in the case where just one of the parties was expanded, it is enough to impose $\Gamma^{(1)}\geq 0$ (see Appendix \ref{cosica}).

\subsection{Expanded bodies in dimension-bound Bell scenarios}
\label{exp}

\begin{figure}
  \centering
  \includegraphics[width=8.5 cm]{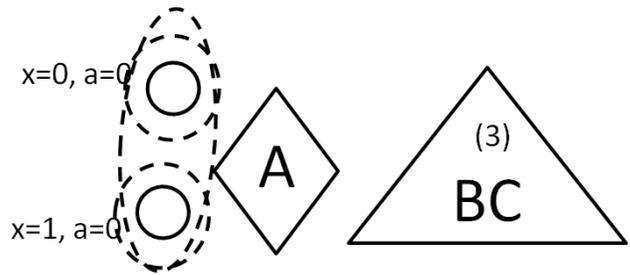}
  \caption{A possible relaxation to bound the convex hull of eq. (\ref{original}) when Alice's measurements are not trusted. Here just Bob and Charlie have expanded bodies, represented by a triangle. The number $(3)$ indicates the order of the moment relaxation.}
  \label{hybrid}
\end{figure}

Consider a tripartite Bell scenario where the local dimension of one of the parties is bounded: the situation is similar to that in the previous section, i.e., eq. (\ref{original}) holds. This time, however, we ignore the mathematical expression of Alice's measurement operators $\{\Pi^x_a\}$. Our solution is, of course, to combine the two previous methods. Figure \ref{hybrid} shows a diagrammatic representation of a possible relaxation for this problem.

This diagram has to be understood as follows: the triangle represents the Hilbert space defined by $\mbox{span}(\ket{s}:|s|\leq 3)$ in eq. (\ref{expansion}); the rest of the figures represent the first Hilbert space in the same expression. This first Hilbert space can be expressed as a tensor product of three Hilbert spaces (Alice's head and her two legs). Note as well that the positivity of the partial transpose of the matrix (\ref{expansion}) with respect to three different subsets of Alice's legs is being enforced. Further -better- relaxations are attained by increasing the order $n$ of the expansion, and the length of Alice's legs.

In such a general case, probabilities are extracted from the main matrix via the formula:

\be
P(a,b,c|x,y,z)=\tr\{(M^x_a\otimes\ket{t}\bra{s})\Gamma^{(n)}\},
\ee

\noindent where $M^x_a$ are defined as in (\ref{pseudo_meas}), and $s,t$ are any two sequences such that $t^\dagger s=E^y_bF^z_c$.

It can be shown (see Appendix \ref{hybrid_ap}) that the above method converges to the convex hull of the set of all distributions of the form

\be
P(a,b,c|x,y,z)=\tr\{(\Pi^x_a\otimes E^y_bF^z_c)\rho\},
\label{commut}
\ee

\noindent with $\Pi^x_a$ ($E^y_b,F^z_c$) acting over a $d$-dimensional (finite or infinite dimensional) Hilbert space, and $[E^y_b,F^z_c]=0$.

When $E^y_b,F^z_c$ act over a finite dimensional Hilbert space, the above expression can be proven equivalent to eq. (\ref{original}) \cite{scholz}, i.e., one can identify commutativity and tensor products. In the infinite dimensional case, though, this is no longer the case, and the existence of a tensor representation for (\ref{commut}) relies on the validity of Kirchberg's conjecture, a major open problem in mathematics \cite{connes1,connes2}. It shall be noted that this technical limitation, already present in the NPA hierarchy \cite{NPA2}, only concerns the \emph{convergence} of the SDP schemes presented in this section. That is, independently of whether Kirchberg's conjecture is true or not, the algorithms proposed above constitute a rigorous relaxation of the original tripartite characterization problem.

%That is, modulo Connes embedding conjecture \cite{connes1,connes2}, this hierarchy of semidefinite programs converges to eq. (\ref{original}).

\begin{remark}
Suppose that we wish to characterize the set of bipartite distributions

\be
P(a,b|x,y)=\tr(\Pi^x_a\otimes E^y_b \rho_{AB}),
\label{bip_prob}
\ee

\noindent with $\{\Pi^x_a,E^y_b\}$ acting over $B(\C^d)$ but otherwise unknown. From the Schmidt decomposition, this scenario can be seen equivalent to just limiting Alice's operators to act over $B(\C^d)$, while allowing Bob's operators to access Hilbert spaces of arbitrarily high (or even infinite) dimension. Hence we can expand Bob's body to the first order while assigning head and legs to Alice. From Appendix \ref{cosica}, it follows that expanding Bob's head to higher orders will not improve the approximation; convergence to (\ref{bip_prob}) is thus achieved simply by increasing the length of Alice's legs. Note that the size of the corresponding generalized moment matrix is still exponential in Alice's number of measurements, but linear in Bob's. With this trick, the 4422 scenario, as well as others of the form $4m22$, can therefore be optimized in a normal computer.

\end{remark}

\section{Application examples}

%\textbf{TO FILL}

As an application of the techniques developed in the preceding sections, we provide several examples. First we discuss dimension witnesses for the 2-party scenario. These witnesses are actually Bell-type inequalities whose certain violation gives a lower bound on the dimension of the Hilbert space. We first apply the entanglement based method of Sec.~\ref{bip} for witnessing dimension in two-party systems using three-setting dichotomic Bell inequalities. Then we move to more demanding Bell inequalities with Alice having four dichotomic settings and Bob having up to twelve dichotomic settings by using the method of Sec.~\ref{tow}. Next we discuss the multipartite case by fixing the local Hilbert space of one of the parties to be two dimensional, but we do not impose any bound on the dimension of the rest of the parties. This hybrid scenario will allow us to certify true three-dimensional entanglement in a device independent manner. For this sake, we make use of a three-party Bell inequality having three dichotomic settings per party, which turns out to be a minimal construction. Finally, it is demonstrated that our technique is also suitable to certify entangled measurements in finite dimensional Hilbert spaces in a rigorous way.

In the following computations we used the MATLAB package YALMIP \cite{yalmip} and the SDP solvers SeDuMi \cite{sedumi}, CSDP \cite{csdp}, SDPLR \cite{sdplr} and SDPNAL \cite{sdpnal}.

\subsection{Two-party dimension witnesses}

\subsubsection{A family of three setting dimension witnesses}

Let us consider the tilted version of the I3322 inequality. This one-parameter family of inequalities $I_{3322}(\eta)\leq 0$ is parametrized by $1/3\le\eta\le 1$, and we refer the reader to the references \cite{tilted1,tilted2} for the explicit form of this family. Strong numerical evidence shows \cite{tilted2} that this inequality cannot be violated by conducting measurements on qubits if $\eta\leq 0.428$. Using the technique of Section~\ref{bip}, we show that the limit is indeed $\eta\simeq 0.428$ subject to numerical precision of the SDP solver SeDuMi \cite{sedumi}. In Table~I we present the two-qubit maximum results for various $\eta$ values. In particular, the lower bound value arises from a see-saw iteration procedure \cite{seesaw}, where all respective measurements turn out to be on the X-Z plane. The upper bound value, on the other hand, is due to the SDP technique of Sec.~\ref{bip}. Note that by $\eta\simeq0.429$ the SDP upper bound value becomes comparable with the precision of our SDP solver ($\sim 10^{-9}$). As it can be observed, the LB and UB values are in good agreement for $\eta\ge0.45$. The complexity of the SDP problem can be characterized by the number of constraints involved and the dimension of the underlying semidefinite matrix. In our particular case, the respective numbers are 2080 and 1027, and solving the SDP problem took about 1 minute on a desktop PC.

\begin{table}
\label{i3322eta}
\begin{center}\begin{tabular}{|c|c|c|}
\hline
  $\eta$ & Lower bound & Upper bound\\
  \hline
  $1$ & 0.25000 & 0.25000  \\
  $0.8$ & 0.14331 & 0.14331 \\
  $0.6$ & 0.03910 & 0.03910 \\
  $0.5$ & 0.00608 & 0.00608 \\
  $0.45$ & $2.8014\times 10^{-4}$ & $2.8015\times 10^{-4}$\\
  $0.44$ & $4.8213\times 10^{-5}$ & $5.8207\times 10^{-5}$\\
  $0.43$ & $1.0764\times 10^{-7}$ & $8.7542\times 10^{-7}$\\
  $0.429$ & $2.9466\times 10^{-9}$ & $5.9880\times 10^{-8}$\\
  $0.428$ & $\sim10^{-17}$ &  $3.7484\times 10^{-9}$\\
  \hline
\end{tabular}
\caption{Lower and upper bounds on the violation of the
$I_{3322}(\eta)$ inequality in the two-qubit Hilbert space. The local bound is equal to 0 for any $\eta$ displayed.}
\end{center}
\end{table}

\subsubsection{Four setting dimension witnesses}

The technique presented in Sec.~\ref{tow} is computationally cheaper than the one of Sec.~\ref{bip} used previously for three setting inequalities. So let us utilize this more powerful technique to construct dimension witnesses with four measurement settings per party.
Firstly consider a four setting tight Bell inequality, which is the $N=4$ member of the INN22 family \cite{CG}.
Here a qubit lower bound is given by $0.25$, when Bob measures a rank-0 projector in one of his settings (and the rest of the measurements are rank-1 projectors). Note that a rank-0 projector accounts for a never-occurring outcome of a measurement. In the following, we will call such measurements \emph{degenerate}. This value of $0.25$ could not be overcome using the see-saw variational technique. The qubit upper bound due to our SDP algorithm is given by $0.26548$, whereas the maximum overall quantum value certified by the NPA hierarchy is $0.28786$, which is attainable with real-valued qutrit systems \cite{PVfull}. Hence, a Bell violation bigger than $0.26548$ serves as a dimension witness, signaling the presence of qutrit systems. In the present case, the number of constraints involved in the SDP problem is 3241 and the dimension of the semidefinite matrix is 883. Our desktop PC required about 15 minutes to solve the problem.

Another inequality, which is not tight but despite its simplicity gives a dimension witness with relatively good noise tolerance, is the following one:
\begin{align}
I_{4,4}=& E^A_1 + E_{1,1}+E_{1,2}+E_{2,1}-E_{2,2} \nonumber\\
&+E_{3,3}+E_{3,4}+E_{4,3}-E_{4,4}\le 5,
\end{align}
where the correlator $E_{x,y}$ between measurement $x$ by Alice and measurement $y$ by Bob is defined as $E_{x,y}=P(a=b|x,y)-P(a\neq b|x,y)$, and $E^A_x$ denotes the single-party marginal of Alice's $x$-th measurement setting.
Notice that the inequality is composed by a CHSH inequality (for settings 3 and 4) and a tilted CHSH inequality (for settings 1 and 2). An upper bound is given by adding up the maximum quantum value of these two Bell expressions \cite{AMP},
$Q=2\sqrt 2 + \sqrt 10\simeq5.9907$. This bound can in fact be saturated by a 2-ququart system, by tensoring a 2-qubit singlet state with a 2-qubit partially entangled state. However, if we fix dimension two for the Hilbert space of both parties, we expect not to attain the overall quantum maximum. Indeed, numerical evidence shows that for qubits the limit is $5.8310$, whereas the upper bound using the expanded bodies technique of Section~\ref{exp} is given by $5.8515$. Hence, a value bigger than $5.8515$ certifies three-dimensional systems. We tried to increase the order of the expansion in order to get even better upper bounds in both above cases, but unfortunately the SDP problem was not feasible using the solvers SeDuMi \cite{sedumi} or CSDP \cite{csdp} on a normal desktop computer.

\subsubsection{Correlation type dimension witnesses}

We investigate the qubit bound of correlation type Bell inequalities, where Alice has four and Bob has up to twelve dichotomic measurement settings. We consider the following linear functions of correlators $E_{x,y}$,
\begin{equation}
I_{m_A,m_B} = \sum_{x=1}^{m_A}\sum_{y=1}^{m_B}{M_{x,y}E_{x,y}}\le L,
\end{equation}
where $m_A$ and $m_B$ are the number of settings on Alice and Bob's side, respectively. Hence, $I_{m_A,m_B}$ defines an ($m_A$,$m_B$) setting correlation type Bell inequality, where $L$ denotes the local bound. Let's take three such Bell inequalities, defined by the coefficient matrices $M$ as follows,

\begin{equation}\label{i47}
M_{4,7} = \left(
\begin{tabular}{ccccccc}
1 & 1  & 1 & 1 & 0 & 0 & 0\\
1 & -1 & 0 & 0 & 1 & 1 & 0\\
1 & 0  & -1 & 0 & -1 & 0 & 1\\
1 & 0  & 0 & -1 & 0 & -1 & -1\\
\end{tabular}
\right),
\end{equation}
and
\begin{equation}\label{i48}
M_{4,8} = \left(
\begin{tabular}{cccccccc}
1 & 1   & 1  &  1 &  1 &  1 &  1 &  1\\
1 & 1   & 1  &  1 & -1 & -1 & -1 & -1\\
1 & 1   & -1 & -1 &  1 &  1 & -1 & -1\\
1 & -1  & 1  & -1 &  1 & -1 &  1 & -1\\
\end{tabular}
\right),
\end{equation}
and
\begin{equation}\label{i412}
M_{4,12} = \left(
\begin{tabular}{cccccccccccc}
1 & 1  & 1 & 1  & 1 & 1  & 0 & 0 & 0 & 0  & 0 & 0\\
1 & -1 & 0 & 0  & 0 & 0  & 1 & 1 & 1 & 1  & 0 & 0\\
0 & 0  & 1 & -1 & 0 & 0  & 1 & -1 & 0 & 0 & 1 & 1\\
0 & 0  & 0 & 0  & 1 & -1 & 0 & 0 & 1 & -1 & 1 & -1\\
\end{tabular}
\right).
\end{equation}
The local bound of the corresponding inequalities $I_{4,7}$, $I_{4,8}$ and $I_{4,12}$ are given by 8,12, and 12, respectively. Note that all above Bell inequalities are members of a larger family~\cite{VPcorr}. In particular, $I_{4,8}$ is a straightforward generalization of Gisin's elegant inequality~\cite{questions}.

\begin{figure}
  \centering
  \includegraphics[width=7.5 cm]{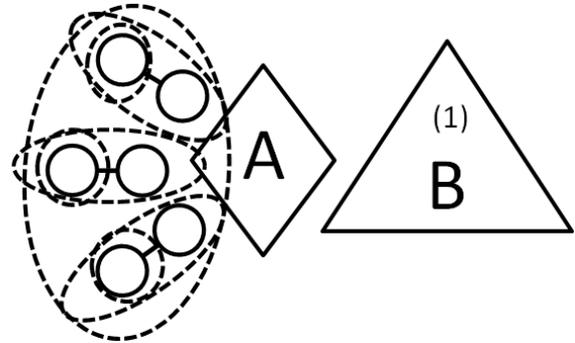}
  \caption{Pictorial representation of the second relaxation used to compute the maximal violation of $M_{4,12}$ in qubit systems.}
  \label{I412}
\end{figure}

Applying the method of Sec.~\ref{exp} for the case of two parties, we get the two-qubit upper bounds summarized in Table~II. As a comparison, the qubit lower bound and ququart maximum values are also given. According to the table, each three inequalities serve as dimension witnesses. Note, however, that there are small gaps between the upper and lower bounds obtained. Hence we programmed a higher relaxation, depicted in Figure \ref{I412}, of the method of Sec.~\ref{exp} to bound the value of $I_{4,12}$, which allowed us to close the gap. To implement this second order relaxation, we used a memory-enhanced desktop and the SDP solver SDPNAL \cite{sdpnal}. This case was the most demanding among all the studied examples from a computational point of view. Here the number of constraints was 1385281 and the dimension of the underlying semidefinite matrix was 13312 and took about 13 hours for our computer to solve the problem.

\begin{table}
\label{corr_res}
\begin{center}\begin{tabular}{|c|c|c|c|}
\hline
  $$ & Qubit lower bound & Qubit upper bound & Ququart\\
  \hline
  $I_{4,7}$ & 10.4995 & 10.5102 & 10.5830 \\
  $I_{4,8}$ & 15.4548 & 15.7753 & 16\\
  $I_{4,12}$ & 16.7262 & 16.7645 (16.7262) & 16.9706 \\
  \hline
\end{tabular}
\caption{Qubit lower/upper bounds on the violation of the
$I_{m_A,m_B}$ inequalities defined by Eqs.~(\ref{i47},\ref{i48},\ref{i412}) computed using see-saw iteration/derived from our construction. The number between brackets in the second column corresponds to the SDP relaxation of Figure \ref{I412}. The ququart value defines the overall quantum maximum given by the zeroth level of the NPA hierarchy.}
\end{center}
\end{table}

As a side note, let us mention that in the present case of correlation type Bell inequalities, it was enough to consider rank-1 projective  measurements (i.e., no need to take into account degenerate measurements), since this type of inequalities is known to be maximized in the two-qubit space by using rank-1 projective measurements~\cite{AGT}.

\subsection{Genuine tripartite higher dimensional entanglement}

Let us consider a three party three setting Bell inequality, which is invariant under any permutations of the three parties and it has the peculiarity that it consists of only 2-party correlation terms, which latter is usually an advantage in experimental situations. The inequality is as follows:
\begin{align}
\label{I333}
I_{3,3,3}=&\text{sym}\{-P(A_1)-2P(A_3)+P(A_1,B_1)\nonumber\\
&-2P(A_2,B_2)-2P(A_3,B_3)-P(A_1,B_2)\nonumber\\
&+P(A_1,B_3)+2P(A_2,B_3)\}\le 0,
\end{align}
where $\text{sym}\{X\}$ denotes that all terms in the expression $X$ have to be symmetrized with respect to all permutations of the parties, and we used the simplified notation $P(A_x,B_y,C_z)=p(0,0,0|x,y,z)$. On one side, we computed lower bound values arising from the heuristic see-saw search for different dimensionalities of the parties, $2\times2\times2$, $2\times3\times3$ and $3\times3\times3$. Note that the cases $2\times d\times d$, $d\times 2\times d$, $d\times d\times 2$ refer to the same situation, because the inequality~(\ref{I333}) is fully symmetric. Therefore, it is enough to perform optimization in one of the cases, say, $2\times d\times d$, where dimension $d\ge2$. On the other side, we give the upper bound value for the case of $2\times\infty\times\infty$ (that is, when Alice acts on qubits, and the other parties have no restriction on the dimension). Due to the symmetry of the inequality, the same upper bound applies to the $\infty\times2\times\infty$ and $\infty\times\infty\times2$ situations, as well.

In the present case, we have to take into account degenerate measurements (either rank-0 or rank-2 projective measurements) on Alice's side, in which case the inequality~(\ref{I333}) reduces to a two setting inequality on Alice's side, hence Alice's qubit state space suffices to obtain maximum quantum violation \cite{Masanes}. That is, when we compute an upper bound on $2\times\infty\times\infty$ in the degenerate case, we can use the dimension unrestricted case of the NPA method \cite{NPA}.
Table~III summarizes the results obtained. By eye inspection, both upper bounds are saturated, hence they are tight (up to numerical precision). Hence, any Bell violation of $I_{3,3,3}$ bigger than $0.1786897$ witnesses in a device-independent way that the underlying state $\rho_{ABC}$, not only has Schmidt number vector $(3,3,3)$ \cite{marcus}, but also that any pure state decomposition of $\rho_{ABC}$ contains at least one state $\sigma_{ABC}=\proj{\psi}$ such that $\mbox{rank}(\sigma_A),\mbox{rank}(\sigma_B),\mbox{rank}(\sigma_C)\geq 3$. To illustrate the power of this Bell inequality, let us pick the following state
\begin{equation}
\label{state3}
\ket{\Psi}=\cos\alpha\ket{\psi}+\sin\alpha\ket{222},
\end{equation}
with $\alpha=0.2519038$, where $\ket{\psi}=(\ket{012}+\ket{021}+\ket{102}+\ket{120}+\ket{201}+\ket{210})/\sqrt 6$ is the fully (bosonic) symmetric 3-qutrit state. By optimizing over the measurement angles in the X-Z plane, we get the quantum value $Q=0.1841287$. Since this value is clearly bigger than the threshold $0.1786897$, we can argue device independently that the above state~(\ref{state3}) is genuinely three-dimensional.

%Note also that the measurement angles are not symmetric in terms of the parties, though both the state and the Bell functional is symmetric with respect to party exchange.

\begin{table}
\label{GTE_res}
\begin{center}\begin{tabular}{|c|c|c|c|c|}
\hline
  $$ & LB & LB & UB & LB\\
  $$ & (222) & (233) & $(2\infty\infty)$ & (333)\\
  \hline
  No-deg & 0.0443484 & 0.1783946 & 0.1783946 & 0.1962852\\
  Deg & 0.1783946 & 0.1786897 & 0.1786897 & $$\\
  \hline
\end{tabular}
\caption{Qubit lower/upper bounds for different local dimensions on the violation of the
$I_{3,3,3}$ inequality computed using see-saw search/SDP computation. Qutrit value $(333)$ is the overall quantum maximum as certified by the NPA hierarchy. The upper bound value for the non-degenerate case (denoted by No-deg) was computed using the technique of Sec.~\ref{exp}, whereas the upper bound value for the degenerate case (denoted by Deg) was obtained by the NPA hierarchy. Abbrevation LB/UB refers to lower/upper bound.}
\end{center}
\end{table}

\subsection{Entangled measurements in two-qubit Hilbert spaces}
\label{ent_meas}

\begin{figure}
  \centering
  \includegraphics[width=8.5 cm]{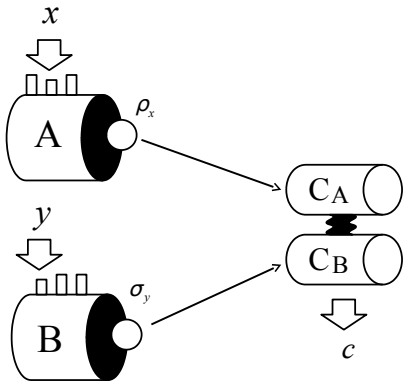}
  \caption{Picturing the scenario of entangled measurements in bounded Hilbert spaces.}
  \label{entmeas}
\end{figure}

Let us consider the following scenario, pictured in Fig~\ref{entmeas}: two separated parties, Alice and Bob, have each a preparation device which prepares unknown qubit states out of 3 possible respective states $\rho_x$ and $\sigma_y$. These states are sent to Charlie's two distinct ports $C_A$ and $C_B$, who in turn interacts with the received states and announces a bit $c$. The experiment is described by a set of conditional probabilities $P(c|x,y)=\tr\left(\rho_x\otimes\sigma_y M_c\right)$, where $M_c, c=0,1$ denote Charlie's POVM elements.

Depending on the form of $M_c$ one can distinguish between different scenarios. In case of \emph{unentangled measurements}, each of the
POVM elements $M_c$ is a separable operator. Moreover, it is known that a subclass of this class corresponds to \emph{LOCC measurements}, in which case $M_c$ is associated with a sequence of measurements on $C_A$ and $C_B$ ports, with each measurement depending on the outcomes of earlier measurements. On the other hand, in case of \emph{general measurements}, the measurement operators in quantum mechanics are only limited by positivity and normalization, and they can be well entangled. For instance, Bell state measurements belong to this class.

We consider the following witness, introduced in~\cite{VN}:
\begin{equation}
W = - P_{11} - P_{12} + P_{13} + P_{21} + P_{23} + P_{31} - P_{32} - P_{33},
\end{equation}
where we identify $P_{xy}=P(0|x,y)$. Using a see-saw type iteration, we obtained the bound $w_{gen}=2.5$ for general measurements and $w_{unent}=(2+3\sqrt 6)/4\simeq 2.3371$ \cite{VN}. Note, however, that due to the heuristic nature of the see-saw type search, these bounds are not rigorous, they constitute only a lower bound to the problem. On the other hand, adapting the technique of Section~\ref{bip} to the present case, we get an upper bound of 2.506 for $w_{gen}$ (in this case, the solver SDPLR was used \cite{sdplr}). In the unentangled case, we identify separable measurements with rank-2 projectors, which may not be justified in general. However, modulo this condition, we get the upper bound of 2.3371 for $w_{unent}$. In the latter case, rank-2 projective measurements are composed by the sum of two orthogonal rank-1 projectors. Then, we have to define a leg for each rank-1 projector, and impose that the legs are orthogonal, as described in Sec~\ref{hig}.

Note that the result $w_{gen}<2.506$ allows us to turn around the problem. Namely, suppose that there is no dimensionality constraint on Alice and Bob's emitted states $\rho_x$ and $\sigma_y$. Then the inequality $w_{gen}\le 2.506$ may work as a dimension witness: its violation guarantees that at least qutrits had to be prepared by Alice or Bob (or by both parties).

\section{Conclusion}
In this paper we have studied the problem of bounding the strength of quantum nonlocality under local dimension constraints. By relating finite dimensional quantum correlations to the separability problem, we have managed to exploit existing entanglement detection criteria to devise hierarchies of SDP relaxations for the characterization of quantum nonlocality in multipartite scenarios with a promise on the local dimensions of the parties involved. The first relaxations of our method were applied successfully to upper bound the maximal violation of several bipartite Bell inequalities in qubits. The relatively small memory resources required to implement our method allowed us to investigate with a normal desktop bipartite Bell scenarios with 4 settings on one side and 12 on the other. Although it was not always possible to close the gap between our upper bounds and the corresponding lower bounds obtained via variational methods, our SDP relaxations output results below the quantum maximum in all cases considered. Moving on to tripartite scenarios, we applied the method to identify a tripartite Bell inequality that cannot be violated maximally if any one of the parties holds a qubit. This inequality can hence be used to certify device-independently that a tripartite quantum state has genuine three dimensional entanglement \cite{marcus}. Finally, we applied the hierarchy to certify entangling measurements in two-qubit Hilbert spaces, as in \cite{VN}.

The reader may have noted that, no matter the dimension of the local Hilbert space, all our examples involved dichotomic measurement operators. The reason is that extremal dichotomic measurements are known to be projective, and so they admit a simple representation in terms of legs. The characterization of many-outcome POVMs is, however, not so straightforward. In principle any POVM can be expressed as a projective measurement in a higher dimensional Hilbert space, and so our method can be adapted, via dimension enhancement, to Bell scenarios involving more than two outcomes. However, the known bounds on the minimal dimension required for arbitrary POVMs are high enough as to make our method impractical in a normal computer, see \cite{chen}. It is an open question whether extreme many-outcome POVMs require considerably less dimension resources, like in the qubit case \cite{dariano} or, more generally, whether the description of extremal POVMs can be simplified to the point of making our new method feasible for such Bell scenarios.

\section*{Acknowledgements}
T.V. acknowledges financial support from the J\'anos Bolyai
Programme of the Hungarian Academy of Sciences and form the
Hungarian National Research Fund OTKA (PD101461). The publication
was supported by the T\'AMOP-4.2.2.C-11/1/KONV-2012-0001 project.
The project has been supported by the European Union, co-financed
by the European Social Fund.

\begin{appendix}
\section{Convergence of expanded bodies}
\label{cosica}

In general, a positive semidefinite operator $\Gamma^{(n)}$ of the form (\ref{expansion}) will not possess a \emph{moment representation}, i.e., there will not exist a state $\rho$ and projector operators satisfying

\be
E^y_bE^y_{b'}= F^z_cF^z_{c'}=0,
\label{orth}
\ee

\noindent for $b\not=b', c\not=c'$, and

\be
[E^y_c,F^z_c]=0,
\label{commute}
\ee

\noindent such that $c^{k,j}_s\equiv\tr(\ket{j}\bra{k}\otimes s\rho)$.

One can, however, prove the following result.

\begin{theo}
Let $\Gamma^{(n)}$ be a positive semidefinite matrix of the form (\ref{expansion}). Then, there exist a finite-dimensional Hilbert space $\H$, a normalized state $\rho\in B(\C^d\otimes \H)$ and projector operators $\{\hat{E}^y_b, \hat{F}^z_c\}\subset B(\H)$ satisfying (\ref{orth}) such that

\be
c^{k,j}_s=\tr\{\rho(\ket{j}\bra{k}\otimes \hat{s})\},
\label{casi_rep}
\ee

\noindent for any sequence $\hat{s}$ of the operators $\{\hat{E}^y_b, \hat{F}^z_c\}$ with $|s|\leq 2n$.

\end{theo}

Note that, due to the structure of the coefficients $\{c^{k,j}_s\}$, even though the commutator $[E^y_c,F^z_c]$ may be different from zero, the identity

\be
\tr\{\rho(\ket{j}\bra{k}\otimes s\hat{E}^y_b\hat{F}^z_c\tilde{s})\}=\tr\{\rho(\ket{j}\bra{k}\otimes s\hat{F}^z_c\hat{E}^y_b\tilde{s})\}
\ee

\noindent must hold as long as $|sE^y_bF^z_c\tilde{s}|\leq 2n$, since both operator products are associated to the same `logical' sequence.

Also notice that, if only one party, say Bob, was expanded, eq. (\ref{casi_rep}) implies that we achieve convergence with $n=1$.

\begin{proof}

The condition $\Gamma^{(n)}\geq 0$ implies \cite{horn} that

\be
c^{k,j}_{t^\dagger s}=\Gamma^{(n)}_{(k,s),(j,t)}=\braket{\psi^j_t}{\psi^k_s},
\ee

\noindent for some collection of vectors $\{\ket{\psi^k_s}\}$. Here, as in the main text, the variables $s,t$ are used to represent operator products; $k,j$, natural numbers ranging from $1$ to $d$. With a slight abuse of notation, if $s$ denotes a null sequence, the corresponding coefficient $c^{k,j}_s$ will be taken equal to zero.

Now, define the vector

\be
\ket{\phi}\equiv\sum_k\ket{k}\ket{\psi^k_{\id}}.
\ee

\noindent It is immediate that this vector is normalized. Indeed, note that

\be
\braket{\phi}{\phi}=\sum_k\braket{\psi^k_\id}{\psi^k_\id}=\sum_k c^{k,k}_{\id}=1.
\ee

\noindent We will hence identify $\ket{\phi}$ with the normalized state in the theorem, i.e., $\rho=\proj{\phi}$.

Now, define the subspaces

\begin{eqnarray}
&\H^y_b=\mbox{span}\{\ket{\psi^k_s}:s=E^y_b\tilde{s},k=0,...,d-1\}, \nonumber\\
&\H^z_c=\mbox{span}\{\ket{\psi^k_s}:s=F^z_c\tilde{s},k=0,...,d-1\}.
\end{eqnarray}

\noindent For $b\not=b'$, The fact that $0=c^{j,k}_0=\braket{\psi^j_{E^y_{b'}s}}{\psi^k_{E^y_b\tilde{s}}}$ implies that $\H^y_b\perp\H^y_{b'}$, and likewise we have that $\H^z_c\perp\H^z_{c'}$, for $c\not=c'$. It follows that the projectors

\be
\hat{E}^y_b\equiv \mbox{proj}(\H_b^y),\hat{F}^z_c\equiv \mbox{proj}(\H_c^z)
\ee

\noindent satisfy

\be
\hat{E}^y_b\hat{E}^y_{b'}=\delta_{bb'}\hat{E}^y_b,\hat{F}^z_c\hat{F}^z_{c'}=\delta_{cc'}\hat{F}^z_c.
\ee

Let us explore how these operators act over the vectors $\{\ket{\psi^k_s}\}$. We have that

\be
\hat{E}^y_b\ket{\psi^k_s}=\hat{E}^y_b\sum_{b'\not=\tilde{b}}\ket{\psi^k_{E^y_{b'}s}}+\hat{E}^y_b\ket{\mbox{rest}}
\label{chicha}
\ee

\noindent with

\be
\ket{\mbox{rest}}=\ket{\psi^k_s}-\sum_{b'\not=\tilde{b}}\ket{\psi^k_{E^y_{b'}s}}
\ee

\noindent (we remind the reader that $\tilde{b}$ represents the measurement outcome not included in the expansion of Bob's body).

Due to the orthogonality relations $\H^y_b\perp\H^y_{b'}$, for $b\not=b'$, the first term on the right hand side of eq. (\ref{chicha}) is $\ket{\psi_{E^y_bs}^k}$. As for the second term, notice that

\be
\braket{\psi^j_{E^y_bt}}{\mbox{rest}}=c^{j,k}_{t^\dagger E^y_bs}-c^{j,k}_{t^\dagger E^y_bs}=0.
\ee

\noindent It follows that $\hat{E}^y_b\ket{\mbox{rest}}=0$. Putting all together, we have that

\be
\hat{E}^y_b\ket{\psi^k_s}=\ket{\psi^k_{E^y_bs}},
\ee

\noindent and, similarly,

\be
\hat{F}^z_c\ket{\psi^k_s}=\ket{\psi^k_{F^z_cs}}.
\ee

\noindent It follows by induction that, for any sequence $\hat{s}$ of the operators $\{\hat{E}^y_b,\hat{F}^z_c\}$,

\be
\hat{s}\ket{\psi^k_{\id}}=\ket{\psi^k_{s}}.
\ee

Finally, we arrive at

\begin{eqnarray}
\tr\{\rho(\ket{j}\bra{k}\otimes \hat{t}^\dagger\hat{s})\}&&=\bra{\psi^j_\id}\hat{t}^\dagger\hat{s}\ket{\psi^k_\id}=\nonumber\\
&&=\braket{\psi^j_t}{\psi^k_s}=c^{j,k}_{t^\dagger s}.
\end{eqnarray}

\end{proof}

Following the lines of \cite{NPA2}, the convergence of the scheme follows from the fact that, for any sequence of positive semidefinite moment matrices $(\Gamma^{(n)})_n$ such that (\ref{proba_ex}) holds, there exists a set of vectors $\{\ket{\psi^k_s}:k=0,...,d-1\}\subset \H$ which allow (using the same construction as in the previous theorem) to build projector operators $\{E^y_b, F^z_c\}\subset B(\H)$ and a quantum state $\rho\in B(\C^d\otimes \H)$ which satisfy eq. (\ref{original}). The proof is nearly identical to the one in \cite{NPA2}, and so it will not be included in this Appendix.

\section{Convergence of heads, legs and extended bodies}
\label{hybrid_ap}

The purpose of this Appendix is to prove that, for very long legs, the matrix that results when we trace out from $\Gamma^{(n)}$ all circles but one on each of the $L$ legs, the result can be approximated by an expression of the form

\be
\sum_k p_k\bigotimes_{l=1}^L\proj{u^k_l}\otimes \tilde{\Gamma}_k^{(n)},
\label{sep_decomp}
\ee

\noindent where $p_k\geq0$, $\sum p_k=1$ and $\tilde{\Gamma}_k^{(n)}$ is a generalized moment matrix representing Alice's head and Bob and Charlie's expanded bodies. Also, in the above expression, orthogonal legs remain orthogonal. In combination with the results of the previous Appendix, this will show that, taking the limits $\lim_{n\to \infty}(\lim_{N\to\infty})$, the proposed hierarchy achieves convergence.

First, denoting by $\H_L$ the Hilbert space associated to Alice's legs, note that, for any positive semidefinite operator $M\in B(\H_L)$,

\be
\tr_L\{(M_L\otimes \id)\Gamma^{(n)}\}
\ee

\noindent is a positive semidefinite operator of the form (\ref{expansion}), but not necessarily fulfilling the normalization condition (\ref{normal}).

Now, given the symmetric space of $\C^d$, $\H_d^N$, consider the trace-preserving CP map $\Lambda: B(\H_d^N)\to B(\C^d)$ defined by:

\be
\Lambda(\bullet)\equiv\left(\begin{array}{c}N+d-1\\N\end{array}\right)\int \tr(\proj{\phi}^N\bullet)\proj{\phi}d\phi.
\label{formu1}
\ee

\noindent This map was proposed in \cite{power} to study the convergence of the DPS hierarchy \cite{trisep}. In \cite{power}, it was shown that it is equivalent to the partially depolarizing channel:

\be
\Lambda(\bullet)\equiv \frac{N}{N+d}\tr_{N-1}(\bullet)+\frac{d}{N+d}\tr(\bullet)\id_d.
\label{formu2}
\ee

By (\ref{formu1}) it is clear that, applying the map $\Lambda$ to any leg in $\Gamma^{(n)}$, the resulting matrix $\hat{\Gamma}$ is of the form (\ref{sep_decomp}). Orthogonal legs may not remain orthogonal, though. However, for any two orthogonal legs $C,D$, by formula (\ref{formu2}), in the limit of $N\gg 1$, $\tr(V(C,D)\hat{\Gamma})$ tends to zero, thus guaranteeing asymptotic orthogonality. Finally, also by eq. (\ref{formu2}), $\hat{\Gamma}$ can be made arbitrarily close in trace norm to the partial trace of $\Gamma^{(n)}$. Note also that the speed of convergence does not depend on the Hilbert space dimension of the expanded bodies, but on the total trace of $\Gamma^{(n)}$.

Finally, let us remark that maps of the form (\ref{formu1}) converge to the identity channel as $O(d/N)$. In order to improve this rate, one can use a second, more complicated map described also in \cite{power}, which induces an $O((d/N)^2)$ convergence. Beware, though! Such a map can only be applied when the PPT condition has been enforced on $\lceil N/2\rceil$ circles of each of the legs \cite{power}.

\end{appendix}

\end{document}